# The van der Waals-Maxwell phase transition, hidden in Sommerfeld-Dirac hydrogen theory, proves that symmetry in the Coulomb bond is broken

*G. Van Hooydonk, Faculty of sciences, Ghent University, Ghent, Belgium*

Abstract. Left unnoticed for almost a century, 1916 Sommerfeld H theory hides a van der Waals-Maxwell phase transition in the Coulomb lepton-nucleon attraction of ground state H. This classical 19th century symmetry breaking effect, important for CPT, is confirmed by observed H $nS_{1/2}$ and $nP_{1/2}$ series. It proves that trying to produce antihydrogen H̲ with $e^+ + p^- \rightarrow$ H̲ does not make sense. Since hydrogen is the major constituent of the Universe, the energy equilibrium of H̲ antimatter and H matter states in natural hydrogen is in line with the Big-Bang hypothesis.

## I. Introduction

Problems with CPT (charge, parity, time) symmetry for H are not yet solved as proved by ongoing H̲ CPT experiments. The problems go back to 1916, before wave mechanics and before spin was discovered, when Sommerfeld [1] used $2^d$ quantum number $\ell$ and relativity to improve Bohr 1913 H $1/n^2$ theory [2]. A decade later, with wave mechanics and spin, Dirac retrieved Sommerfeld H theory exactly [3]. This Sommerfeld-Dirac (SD) theory was considered absolute at the time. Since it implied that H nS/nP states are degenerate, the theory was flawed by the Lamb-shift [4]. This led to QED with vacuum fluctuations for the H Coulomb field and virtual pairs, which secure the field is $-e^2/(r \pm \delta r)$ rather than straight $-e^2/r$ [5,6]. For Heisenberg [7a], it was a miracle that the same equation was obtained with (Dirac) and without (Sommerfeld) wave mechanics and spin. This theoretical puzzle [7b] may well be the reason why other results of SD theory were overlooked. In fact, we now show that Sommerfeld 1916 H theory [1] gives away a classical 19th century Van der Waals (vdW)-Maxwell [8] phase transition for the H Coulomb field. This vdW curve is confirmed by observed [9] and by QED data [10,11]. Although vdW-like phase transitions (superconductivity, Bose-Einstein condensation, symmetry breaking, chirality, matter-antimatter transitions, order-disorder transitions...[12,13]) follow double well potentials (DWPs), all symmetry breaking effects are similar. The forgotten vdW-transition between 2 H phases has consequences for theory as well as for experiment, as we show below.
The outline is as follows. Section II contains the theory of the SD phase transition of DWP- or Mexican hat-type. Section III shows the theoretical SD and observed vdW-Maxwell curves for both Lyman H nP and nS series. The general discussion in Section IV is followed by applications of broader interest: (i) classical H symmetry breaking (Section V) and (ii) prospects for ongoing CPT experiments and the Big-Bang (Section VI). The conclusion is in Section VII.

## II. Sommerfeld-Dirac phase transition in the H Coulomb field (DWP mode)

With H ground state energy $-E_1 = \tfrac{1}{2}\mu\alpha^2c^2 = \tfrac{1}{2}e^2/r_1$, Bohr energies $E_n$ and terms $T_n$ are

$$E_{n(B)} = E_1/n^2 \text{ and } T_{n(B)} = |E_1|(1-1/n^2) \qquad (1)$$

Bohr assumed thereby that electron-proton Coulomb attraction in atom H

$$V = -e^2/r = -e^- e^+/r \qquad (2)$$

is exactly obeyed from $n=1$ to $n=\infty$. To test (2), we make advantage of the strength of Bohr theory (1), which is that monitoring ground state H is easy with all n H ground states

$$E_{1,n} = n^2 E_n \text{ or } r_{1,n} = \tfrac{1}{2} e^2/(n^2 E_n) \qquad (3)$$

obtained by correcting observed $E_n$ with size scale factor $n^2$. The result can be interpreted as the spectrum of ground state H in the complete n range. In this notation, Bohr's $E_1$ becomes $E_{1,1}$, Bohr length $r_0 = r_{1,\infty}(1+m/M)$. If (2) were valid, $E_{1,n}$ is constant: $E_{1,n} - E_{1,1}$ or $r_{1,n} - r_{1,1}$ should be zero. An analytical test for Coulomb field (2) with (3) is provided by the SD H equation, identical for $nS_{1/2}$ and $nP_{1/2}$, since $j+\tfrac{1}{2}=1$ in either case. For $Z=1$, SD theory gives

$$E_n = \mu c^2 \{[1 + \alpha^2/(\sqrt{(1-\alpha^2)}+n-1)^2]^{-1/2} - 1\} \qquad (4)$$

$$E_{1(B)} = E_{1,1} = \mu c^2 \{[1 + \alpha^2/(1-\alpha^2)]^{-1/2} - 1\} = \mu c^2 \{[1+\alpha^2(1+\alpha^2+\alpha^4\ldots)]^{-1/2} - 1\} \qquad (5)$$

$$T_n = E_{1,1} - E_n = \mu c^2 \{[1+\alpha^2/(\sqrt{(1-\alpha^2)}+n-1)^2]^{-1/2} - [1+\alpha^2/(1-\alpha^2)]^{-1/2}\} \qquad (6)$$

Expanding (4) in $1/n$ leads to

$$E_n = -(\tfrac{1}{2}\mu c^2 \alpha^2/n^2)[1+\alpha^2(1/n-\tfrac{3}{4}/n^2)+\ldots] = (E_{1,\infty}/n^2)[1+\alpha^2(1/n-\tfrac{3}{4}/n^2)+\ldots] \qquad (7)$$

$$E_{1,1} = -\tfrac{1}{2}\mu c^2 \alpha^2 (1+\tfrac{1}{4}\alpha^2+\ldots) = E_{1,\infty}(1+\tfrac{1}{4}\alpha^2+\ldots) \qquad (8)$$

where Rydberg $R = \tfrac{1}{2} m_e c^2 \alpha^2/(hc)$, $-E_{1,\infty} = R_H = R/(1+m_e/m_p)$ and Sommerfeld's $\alpha = e^2/(\hbar c)$.

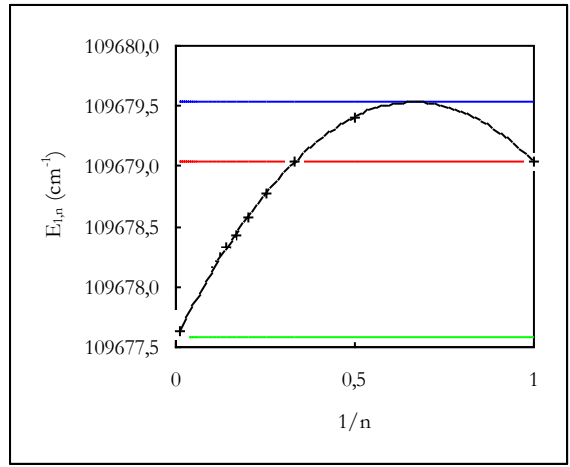

Fig. 1 Parabola (9): $E_{1,n} = n^2 E_n$ versus $1/n$. The 3 asymptotes shown are harmonic $E_{1,3/2} = E_{1,\infty} + 1{,}9$ cm$^{-1}$ for $r_{1,3/2}$ (blue), Bohr's $E_{1,1} = E_{1,\infty} + 1{,}46$ cm$^{-1}$ for $r_{1,1}$ (red) and $E_{SD} = E_{1,\infty}$ for $r_{1,\infty} = r_0$, the Bohr length (green).

SD result (7) provides directly with the spectrum of ground state H $E_{1,n}$ (3)

$$E_{1,n} = n^2 E_n = E_{1,\infty}[1+\alpha^2(1/n-\tfrac{3}{4}/n^2)+\ldots] \qquad (9)$$

which implies that $E_{1,n}$ and $r_{1,n}$ are not constant but obey a parabola in $1/n$. Fig. 1 shows the $E_{1,n}(1/n)$ plot, including asymptotes $E_{1,3/2}$ and $E_{1,\infty}$. Differences with harmonic $E_{1,3/2}$ (blue) and SD asymptote or $E_{1,\infty}$ (green) all have the same sign, respectively + or –. The difference with $E_{1,3/2}$ is a perfect parabola, the square root of which quantifies linear symmetry breaking. Differences with Bohr's $E_{1,1}$ or $r_{1,1}$ are fluctuations (+ and -). SD (3) further implies that the most stable H nP ground state is at n=3/2, not at n=1. The harmonic asymptote at n=3/2 is

$$E_{1,3/2}=E_{1,\infty}(1+\alpha^2/3) \qquad (10)$$

The oscillator part in (9) is a Kratzer potential of type $-a/n+b/n^2$ [14,15], since

$$(1/n-\tfrac{3}{4}/n^2)\equiv[1-(1-1,5/n)^2]/3 \qquad (11)$$

This potential, called after Sommerfeld's assistant Kratzer, gives ¼ for n=1, see (8).

Fig. 1 reveals how symmetry of Coulomb's $-e^2/r_{1,n}$ (2) is broken in ground state spectrum $E_{1,n}$. Although all n ground states are bound, states with large n are less attractive than those with small n. Bohr $E_{1,1}$, in between $E_{1,\infty}$ and $E_{1,3/2}$, is closer to the series limit, which makes it the preferential asymptote to discuss finer details. With $E_{1,1}$, the net effect of SD symmetry breaking (9) on observed terms, attenuated by $1/n^2$, obeys a quartic

$$\Delta_n=E_{1,n}/n^2-E_{1,1}/n^2=E_{1,\infty}[1+\alpha^2(1/n-\tfrac{3}{4}/n^2)+\ldots-(1+\tfrac{1}{4}\alpha^2)]/n^2 \qquad (12)$$

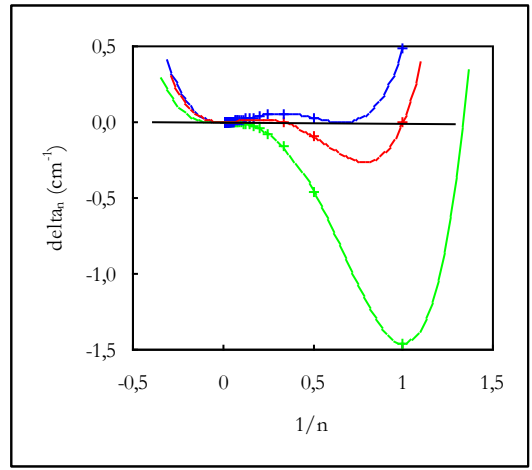

Fig. 2 The 3 DWP curves for SD H $\Delta_n$ (12) versus 1/n with (a) Bohr asymptote $E_{1,1}$ (red), (b) harmonic asymptote $E_{1,3/2}$ (blue) and (c) SD asymptote $E_{1,\infty}$. The straight black line is the linear fit of all data points on the red curve.

In the $\Delta_n(1/n)$ plot of Fig. 2, (12) gives the middle DWP curve. Critical points, available from $d\Delta_n/dn=0$, are n=3+√3=4,7321 and n=3-√3=1,2795 (1/n=0,2113 and 1/n=0,7887). Fig. 2 also shows equivalent differences $\Delta_n$ with harmonic $E_{1,3/2}$ and SD $E_{1,\infty}$, situated respectively above and below the middle curve with series limit $E_{1,1}$. The DWP for harmonic $E_{1,3/2}$ (blue) is symmetric [16], that for Bohr's $E_{1,1}$ (red) is less symmetric. Unless extrapolated, wells are hardly visible with the quartic for SD asymptote $E_{1,\infty}$ (green). Extrapolating branches is possible with (11). The SD quartic, restricted to data points without extrapolation, would show one well instead of two. Below 1/n=0, H does not longer exist, since it is replaced by its dissociation products $e^-$ and $p^+$. For these domains, virtual particle pairs can be invoked. The meaning of the straight line in Fig. 2, obtained by a linear fit for all data points on the middle DWP, is discussed below.

Original 1916 Sommerfeld H theory [1] gives away a DWP (Mexican hat curve), typical for most observed phase transitions at any level or scale (see Introduction).

**III. Sommerfeld-Dirac phase transition in the H Coulomb field (vdW-Maxwell mode)**

*a. $nP_{1/2}$*

Unlike the $\Delta_n(1/n)$ plot in Fig. 2, H symmetry breaking effect (12) can also be discussed with a $\Delta_n(n)$ plot. Since the middle DWP in Fig. 2 is closest to the observed terms and the series limit, the equivalent $\Delta_n(n)$ plot is shown in Fig. 3 (red curve). With n, a classical 19$^{th}$ century vdW-Maxwell curve appears for the SD H phase transition. Since we used n up to 100 by extrapolating (11), a log scale for n is adopted. The straight line, also obtained by a linear fit, is a Maxwell line for coexisting states. Since fitting in n is impossible for (12), the curve shown is an aid to the eye.

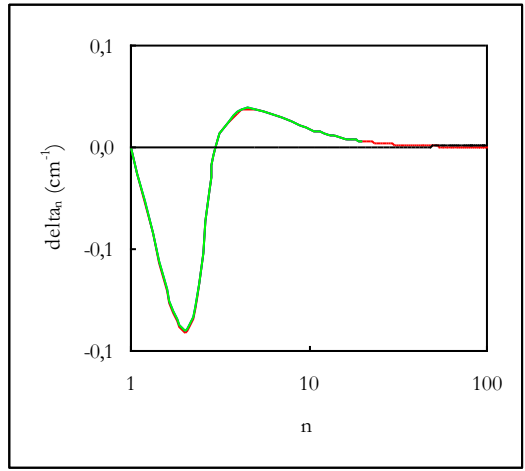

Fig. 3 vdW-Maxwell curve for SD H nP $\Delta_n$ (12) versus n (red), including the
Maxwell line (black). Curves from QED $E_n$ [10] (blue) and from observed $T_n$ [9] (green),
corrected for the 1S-1S Lamb shift. (log scale for n)

This classical thermodynamic result for H can be understood from ideal gas law pv=kT. Here, pv reduces to a 1/r law like Coulomb's: it is a force (1/$r^2$) on surface ($r^2$) multiplied by volume ($r^3$). Like the SD DWP curve (Fig. 2), a SD van der Waals curve equally describes small fluctuations in H ground state Coulomb attraction. With this classical view on a phase transition, it is evident that 2 different ground state phases, H I and H II, must exist.

This theoretical SD vdW curve is fully confirmed by observed [9] and QED data [10]. With average differences smaller than 1 MHz, curves collapse, as illustrated by the hardly visible green and blue curves, extracted from [9] and [10].

*b. $nS_{1/2}$*

SD theory may be exact for $nP_{1/2}$, it fails on $nS_{1/2}$ due to the Lamb shift [4]. nS $E_{1,nS}=n^2 E_{nS}$ in (3) can be calculated from observed [9] and QED data [10,11]. In the nS parabola

$$E_{1,nS}=n^2 E_{nS}=E_{1,\infty}[1+\alpha^2(0,951706/n-0,748199/n^2)] \tag{13}$$

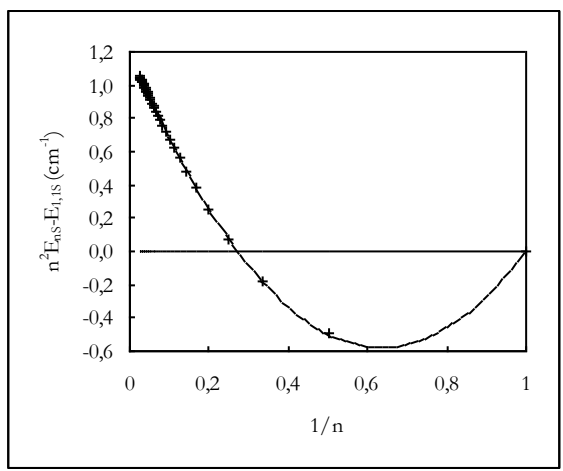

Fig. 4 Parabola (13) for nS ground states $E_{1,nS}$ using QED $T_n$ [11]

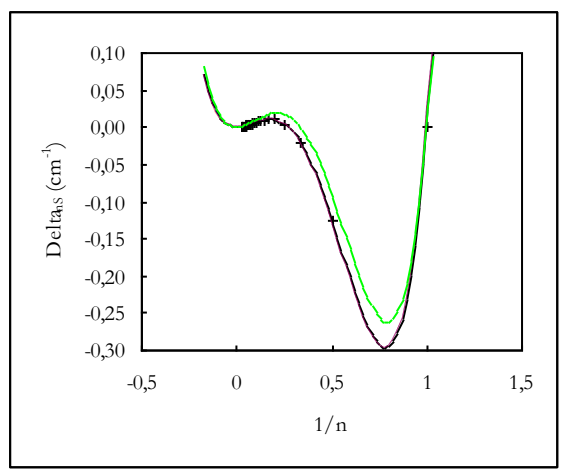

Fig. 5 DWPs for H nS in a $\Delta_{nS}$ (1/n) plot: from QED $T_n$ [11] (black), QED $E_n$ [10] (blue), observed $T_n$ [9] (red). SD DWP for H nP (green) is included for reference

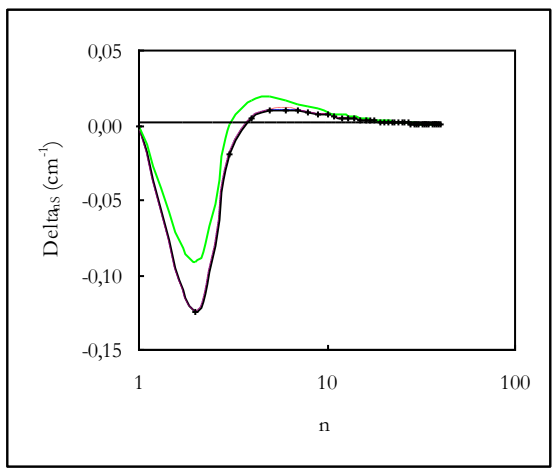

Fig. 6 vdW-Maxwell curves for H nS in a $\Delta_{nS}$ (n) plot (colors as in Fig. 5), including Maxwell line (black) from a linear fit of QED $T_n$ data points [11]. The nP vdW curve is included for reference (log scale for n) based on [11], coefficients differ slightly from those in (9). Fig. 4 shows that the extreme for nS at

$$n=1,572333 \qquad (14)$$

is also slightly different from 1,5 in nP. As above for nP, curves for observed $T_n$ [9] and QED $E_n$ [10] coincide (not shown). Symmetry breaking in observed nS terms obeys

$$\Delta_{nS}=E_{1,nS}/n^2-E_{1,1S}/n^2=E_{1,\infty}[1+\alpha^2(0,951706/n-0,748199/n^2)-(1+0,203506\alpha^2)]/n^2 \quad (15)$$

to be compared with (11) for nP. Critical n=1,282813 is for the maximum, n=5,751999 for the minimum. The $\Delta_{nS}(1/n)$ plot with Bohr's asymptote $E_{1,1S}$ (15) in Fig. 5, gives a DWP or Mexican hat curve and includes that for nP (green). A similar harmonic DWP for H nS is in [16].
As above, the $\Delta_{nS}(n)$ plot in Fig. 6 shows the nS vdW curve. A linear data fit gives Maxwell's conjecture for coexisting H phases I and II. The vdW nP curve (green) is included for reference, as are the curves for observed [9] (red) and QED data [10] (blue). As above, nS curves all collapse. The virtual 1S Lamb shift of 8172,87 MHz or 0,2726 cm$^{-1}$ [17], i.e. the difference between SD nP $E_{1,1}$ and $E_{1,1S}$, would normally separate nS and nP vdW curves. The shift generates slightly different critical points but the general shape of vdW curves is not affected. These analytical details are not discussed here.
Fluctuations in Coulomb attraction for isotopes D and T and hydrogen-like He$^+$, Li$^{++}$... follow similar patterns if suited data for reduced mass and Z are used. For hydrogen-like atoms [10], $\Delta_n$ is larger since it varies with Z$^4$. This extension is, however, beyond the scope of this work.

**IV. Discussion**

The Sommerfeld-Dirac phase transition in the 2-unit charge Coulomb bond in atom H is established without doubt for nS and nP. Terms differ in absolute value due to Lamb shifts but symmetry breaking remains similar. This built-in phase transition in the Coulomb H bond in SD theory can be interpreted either with a DWP (Mexican hat) or with a vdW-Maxwell curve. The classical H vdW-Maxwell variant clearly exposes a phase transition between ground states H I and H II, due to fluctuations in Coulomb attraction –e$^2$/r$_{1,1}$ (3). It could have been detected nearly a century ago, as it is contained in original 1916 Sommerfeld H theory with only two quantum numbers n and ℓ (wave mechanics and spin were not yet known). This theory appeared many years before the roaring 1920s with Schrödinger, Heisenberg, Born…. The Dirac version with wave mechanics and spin is identical to the original Sommerfeld H equation of state [7].
The vdW-Maxwell phase transition can only be obtained with SD theory if (i) the limit on observed data is $E_{1,1}$ as in Bohr theory (1), not the SD limit $E_{1,\infty}$ and (ii) if n instead of 1/n is used as variable. Asymptote and variable choices are critical to find the H SD phase transition, which may explain why it remained unnoticed for so long.
Since Dirac's version consumes all available quantum numbers n, ℓ and s, an extra degree of freedom is required to explain this phase transition. We review some possibilities.

(a) With the SD vdW variant, well established from the 19th century on for macro-systems, H Coulomb symmetry breaking can be caused by a rearrangement of the 2 particles in H. The 2-unit charge Coulomb attraction is insensitive to a permutation of charges in

$$V_{1,n} = -e^-e^+/r_{1,n} \qquad (16)$$

for the H electron-proton Coulomb bond to

$$V_{1,n} = -e^+e^-/r_{1,n} \qquad (17)$$

for an $\underline{H}$ positron-antiproton bond, unknown at the time of Bohr and Sommerfeld. Although SD (5) consumes all available quantum numbers n, ℓ and s=±½, the vdW-curves nevertheless show that Coulomb symmetry (16)-(17) is broken. As expected, the parabola in Fig. 1 reveals that Coulomb symmetry breaking (quantum) for H nP ultimately follows Kratzer's linear ±(1-1,5/n).

(b) Vacuum fluctuations of QED bear on the explanation of the Lamb shift by Bethe [18] (with Bethe logarithms [10]) and Welton [5,6]. Now, vdW fluctuations are understood with 2 phases H I and II, situated at either end of the vdW curve (just like liquid and gas). Critical points are in line with coexisting states. The maximum, where 2 phases coexist, is at n=4,7 for $nP_{½}$ (pending Lamb shifts) and at 5,7 for $nS_{½}$. If H phases I and II were chiral (see below), H is a racemic mixture at the intersection point with the Maxwell line. Virtual pairs are confined to theoretical (not observed) domains 1/n<0 and 1/n>1 as indicated by extrapolated branches in Fig. 2 and 5.

This interpretation of the H spectrum leads to a pre-quantal 1873 vdW equation of state (EOS) [8], originally observed only for macroscopic phase transitions. In reduced form, this EOS is generic and valid for any phase transition at whatever scale, irrespective of the number of particles. Maxwell's thermodynamic rule, obeyed by prototypical quantum system H, is the equivalent of Bohr's hypothesis that Coulomb attraction $V=-e^2/r_{1,n}$ be n invariant (2).

(c) Although it was not the primary intention of SD to expose this phase transition, a Kratzer oscillator accounts for fluctuations in the H Coulomb field. This potential, absent in QED, also provides with a solution for oscillations in bond $H_2$, the prototypical oscillator in nature [19-21].

(d) Modern H CPT experiments (see Section VI) were probably conceived without realizing that Sommerfeld 1916 H theory already suggested chiral/mirror symmetry for H. Its predicted DWPs or Mexican hat curves for $nS_{½}$ and $nP_{½}$ are not only in line with the 1874 Boltzmann quartic [22] but also with Hund quantum DWPs for chiral systems [23]. A chiral option for H is supported by Hund-type DWPs in Fig. 2 and 5 and especially by critical n in the $nS_{½}$ parabola in Fig. 4, close to ½π [24]. H mirror symmetry implies that charge permutation takes place, see (16)-(17). Left-handed H I transforms in right-handed H II, antihydrogen $\underline{H}$ (or vice versa). Chiral behavior implies polarization but measuring (small) variations in H line intensities is difficult. A solution can be provided with polarization dependent wavelength shifts [25].

(e) Schrödinger retrieved Bohr H result (1), i.e. without fluctuations in the Coulomb field: there is no room for an H phase transition in wave mechanics unless used in Dirac's way [3]. Uncertainties (Heisenberg) and probabilities (Born) are in (g-h) below.

(f) Alternatives for chiral behavior are fractal, chaotic [26] and Euclidean H behavior, important for physics as proved recently [27]. If deviations from circular orbits were spiral, a connection with Euclid's $\varphi=½(\sqrt{5}-1)$ [23,24] must exist. A link is provided by the maximum in the $nS_{½}$ parabola in Fig. 4 at n=1,572333 (18). This is closer to Euclidian value [26]

$$2\sqrt{\varphi}=2\sqrt{[½(\sqrt{5}-1]}= 2\sqrt{0,6180334}=1,572303 \ (≈½\pi=1,570796) \qquad (18)$$

than to $½\pi$ for mirror symmetry, as suggested previously [24]. Small differences between mirror and Euclidean behavior may be important for matter-antimatter symmetry [28], see Section VI.

(g) Interpreting the H phase transition as an order-disorder transition brings in uncertainties and, by extension, probabilities and 19$^{th}$ century entropy (if not information entropies [29]).

(h) Two H ground states can be understood either in 2D, whereby radius $r_{1,n}$ is either smaller or larger than $r_{1,1}$, or in 3D. Then, if H were really two sided (chiral) instead of varying in size, it would reduce to an infinitesimally thin coin with radius $r_{1,1}$ but with different sides (heads and tails) according to out-of-plane variations, following $r_{1,n}$. The outcome of a Bernoulli trial with this coin depends on n: for small n, tails would show, heads for large n (or vice versa). Visualizing H energy differences by tossing a unit H coin in the Bernoulli trial is like a *Gedankenexperiment*. If carried out, gravitation comes into play, a major challenge in H̲ research as in AEGIS [31].

(j) Fluctuations in $-e^2/r_{1,n}$ (3) can be ascribed to $e^2$. Here, $\sqrt{(n^2E_n/E_{1,\infty})}≈1+\alpha^2[1-(1-1,5/n)^2]/6$ would imply that a unit charge e in H nP could fluctuate by $\sim R\alpha^2/6$.

**V. Classical symmetry breaking of the H Coulomb field**

H symmetry breaking is understood classically [16] with the division of H mass $m_H$ in 2 particles with mass $m_e$ and $m_P$, which always gives two solutions

$$\text{(a) } m_H=m_e+m_P=m_e+(m_H-m_e) \text{ and (b) } m_H=-m_e+(m_H+m_e) \qquad (19)$$

Bohr reduced H mass $1/\mu_{H(I)}=1/m_e+1/m_P=(1/m_e)(1+m_e/m_P)$ for H phase I

$$\mu_{H(I)}/m_e=1/(1+m_e/m_P)\equiv 1-m_e/m_H \qquad (20)$$

leads to a symmetric companion for H phase II, i.e.

$$\mu_{H(II)}/m_e=1+m_e/m_H=1+m_e/m_P-(m_e/m_P)^2\ldots \qquad (21)$$

The underlying classical equilibrium condition for Bohr H phase I (20), i.e.

$$mr=MR \qquad (22)$$

generates an extra degree of freedom. The charge separation in (16)-(17) becomes [16]

$$r_H=\sqrt{(r^2+R^2-2rR\cos\theta)}=(r+R)\sqrt{[1-2rR(1+\cos\theta)/(r+R)^2]} \qquad (23)$$

The pair of classical ± solutions for states H I and II is

$$r_{H(I)}=r+R \text{ for } \cos\theta=-1 \ (\theta=\pi) \text{ and } r_{H(II)}=r-R \text{ for } \cos\theta=1 \ (\theta=0) \qquad (24)$$

With the Rydberg, states (24) would differ by ±59 cm$^{-1}$. Although this is much larger than the SD spread of α$^2$R$_H$/3≈1,9 cm$^{-1}$ for nP in Fig. 1, observed sinusoidal variations in Fig. 1 and 4 are in favor of (29)-(30). How to accommodate for this discrepancy is dealt with elsewhere. Conceptually, classical symmetry breaking (24) can deal with observed fluctuations in the attractive H Coulomb field. Hence, quantifying H P-symmetry (parity) breaking is in reach.

As mentioned in [24], the (chiral) symmetry breaking ratio χ from (19), (20) or (24) amounts to

$$\chi = (1-m/M)/(1+m/M) \approx 1-2m/M+\ldots = 0,0010886+\ldots \quad (25)$$

To do justice to Bohr and Sommerfeld, there was no reason at the time to distinguish between -e$_e^-$e$_P^+$/r (16) and charge inverted -e$_e^+$e$_P^-$/r (17), both obeying -e$^2$/r. Although (24) can lead to unwanted singularities, it is useful for a classical theory for an H phase transition.

## VI. Prospects of an H vdW phase transition for CPT experiments and the Big Bang

In the mid 1990s, CERN/Fermilab [31, 32] argued that antihydrogen <u>H</u> could be made using the then unsuspected Coulomb attraction -e$^+$p$^-$/r (17) in long range reaction

$$e^+ + p^- \rightarrow \underline{H} \quad (26)$$

Its justification derives from the well-established long range Bohr Coulomb e$^-$p$^+$/r attraction (16)

$$e^- + p^+ \rightarrow H \quad (27)$$

At first sight, (26) seems plausible. At the time, many scientists, including myself [33], believed (32) was justified. Yet, (26) would be sound, if it were not for a Coulomb phase transition in H.

The SD H phase transition in the vdW-Maxwell curve in Fig. 6 for nS proves that long-range reaction (26) for synthesizing <u>H</u> cannot make sense. In fact, observed long-range interaction (27) at the right side of this curve, i.e. for large n (n> 5,7) or for large separation r, applies to Bohr phase H I, i.e. for Coulomb attraction -e$^-$p$^+$/r$_{1,n}$ (16). In this far right, Bohr H phase I is predominant. Moving to the left, i.e. by decreasing r or n, H phase I gradually transforms in H phase II. Hence, H state II can only be reached at very short range, i.e. starting at the far left of the maximum of the SD vdW-Maxwell curve in Fig. 6. Since reaction (26) can only be valid at small r, attempts to produce <u>H</u> with long-range reaction (26) are flawed. Classical and common sense physics learns that the natural order of any 2 phases of the same species (water and steam, both H$_2$O) must not be inverted. Or, H phase II or <u>H</u> can never be synthesized with (32). This may explain the many experimental difficulties with (32). Nearly 20 years after <u>H</u> claims [31,32], the <u>H</u> 1S-2S line, believed to be important for CPT, was not yet measured. Only indirect evidence, based on annihilation signatures, is presented [36-38]. Recent reports [39,40] do not provide with new insights on the internal structure of H or <u>H</u> either but mainly repeat earlier expectations about the outcome of <u>H</u>-experiments.

The SD H phase transition predicts that state 1S-2S for natural H with n 2<5,7 must have a different symmetry than all (Bohr-like) hydrogen terms with n>5,7. Consequently, if term 1S-2S with n=2 applies for an antihydrogen state, it is already available with a precision of order 10 Hz (3.10$^{-10}$ cm$^{-1}$)[41].

As shown above, the SD H phase transition is already clearly visible with data, precise to $10^{-4}$ cm$^{-1}$ as in [9], which places question marks on ongoing H̲ experiments.

To get at finer CPT details, it is essential that spectra of H- and H̲-states be understood in the complete interval 1/n=0 to 1/n=1, a topic on which work is in progress.

Finally, the Standard Model (SM) states that H and H̲ spectra are identical. This is correct in the understanding that the spectrum of natural H exposes hydrogen (matter) states at large n and antihydrogen (antimatter) H̲ states at small n. The fact that the spectrum of natural H gives away matter H- as well as antimatter H̲-states, is important for cosmology. Instead of being absent in the Universe, antimatter-states are in perfect equilibrium with matter states, as proved by the Maxwell coexistence line in H vdW curves. Since hydrogen is the most abundant species in the Universe, our results on matter-antimatter symmetry support the Big-Bang hypothesis [42]. The 3 DWPs in Fig. 2, describing matter-antimatter transitions in H, are similar to 3 DWPs in Fig. 3 of [42], deriving from a quartic Higgs potential. Atom H could well be prototypical for symmetry breaking too, since it is the only natural system available, for which symmetry breaking can be measured with a precision of kHz [41].

**VII. Conclusion**

A classical vdW-Maxwell phase transition in 2 unit charge Coulomb interaction $-e^2/r$ in atom H, hidden in 1916 SD theory but always overlooked, is established beyond doubt. The origin of this forgotten H symmetry breaking effect, given away by vdW- or DWP curves, probably resides in classical physics. Trying to find new physics on the basis of modern H̲ experiments seems useless, since the H vdW curve shows that the long-rang reaction used to synthesize H̲ is not tolerated by nature. The H SD vdW phase transition, predicted a century ago, not only falsifies H̲ experiments but also shows why electrically neutral antimatter is abundant in the Universe.

References

[1] A. Sommerfeld, Ann. Phys. (Berlin) **356** 1 (1916)
[2] N. Bohr, Phil. Mag. **26** 1 (1913)
[3] P.A.M. Dirac, Proc. Roy. Soc. A **128** 610 (1928)
[4] W.E. Lamb Jr. and R.C. Retherford, Phys. Rev. **72** 241 (1947)
[5] M.I. Eides, H. Grotch and V.A. Shelyuto, Phys. Rep. **342** 63 (2001); hep-ph/0002158
[6] T.A. Welton, Phys. Rev. **74** 1157 (1948)
[7] (a) W. Heisenberg, Phys. Blätt. 24 530 (1968); (b) L.C. Biedenharn, Found. Phys. **13** 13 (1983); P. Vickers, Eur. J. Phil. Sci. **2** 1 (2012); Y.I. Granovskii, Phys-Uspekhi **47** 523 (2004); A.A. Stahlofen, Phys. Rev. A **78** 036101 (2008)
[8] J. D. van der Waals, Over de continuïteit van de gas- en vloeistoftoestand, Ph. D. Thesis, Leiden, 1873
[9] R.L. Kelly, J. Phys. Chem. Ref. Data **16** Suppl. 1 (1987)
[10] G.W. Erickson, J. Phys. Chem. Ref. Data **6** 831 (1977)
[11] QED H $T_{nS}$ at http://physics.nist.gov/PhysRefData/HDEL/index.html
[12] Y. Nambu, Rev. Mod. Phys. **81** 1015 (2009)
[13] P.G. De Gennes, Rev. Mod. Phys. **46** 597 (1974)
[14] A. Kratzer, Z. Phys. **3** 289 (1920)
[15] A. Kratzer, Ann. Phys. **67** 127 (1922)
[16] G. Van Hooydonk, Acta Phys. Hung. **19** 385 (2004)

[17] M. Weitz et al., Phys. Rev. A **52** 2664 (1995)
[18] H.A. Bethe, Phys. Rev. **72** 339 (1947)
[19] G. Van Hooydonk, Z. Naturforsch. **64a** 801 (2009); Phys. Rev. Lett. **100** 159301 (2008)
[20] M Toutounji, J. Chem. Theor. Comp. **7** 1804 (2011)
[21] K. J. Oyewumi, Int. J. Theor. Phys. 49 1302 (2010)
[22] L. Boltzmann, Pogg. Ann. Phys. Chem., Jubelband, 128 (1874)
[23] F. Hund, Z. Phys. **43** 805 (1927)
[24] G. Van Hooydonk, Phys. Rev. A **66** 044103 (2002)
[25] G. Van Hooydonk, arxiv:physics/06212141 (2006)
[26] G. Van Hooydonk, http://arxiv.org/abs/0902.1096 (2009)
[27] R. Coldea et al., Science **327** 177 (2010)
[28] G. Van Hooydonk, Eur. Phys. J D **32** 299 (2005)
[29] C.E. Shannon, Bell Syst. Tech. J. **27** 379 (1948); ibidem **27** 623 (1948)
[30] C. Canali et al. Eur. Phys. J. D **65** 499 (2011)
[31] G. Baur et al., Phys. Lett. B **368** 251 (1996)
[32] G. Blanford et al., Phys. Rev. Lett. **80** 3037 (1998)
[35] G. Van Hooydonk, Spectrochim. Acta A **56** 12 (2000)
[36] Y. Enomoto et al., Phys. Rev. Lett. **105** 243401 (2010)
[37] G.B. Andresen et al., Phys. Lett. B **685** 141 (2010)
[38] G.B. Andresen et al., Nature **468** 673 (2010)
[39] C. Amole et al., Nature **483** 439 (2012)
[40] G. Gabrielse et al., Phys. Rev. Lett. **108** 113002 (2012)
[41] T. Udem, R. Holwarth and T.W. Hänsch, Nature **416** 233 (2002)
[42] M. Dine and A. Kusenko, Rev. Mod. Phys. **76** 1 (2003)